\documentstyle[prl,aps,preprint,epsf]{revtex}

\begin{document}
\draft
\title{Using a hybrid superconducting-ferromagnetic 
tip as a magnetic scanning tunneling microscope} 
\bigskip
\author{Denis Feinberg$^*$, Guy Deutscher$^\dagger$}
\address {$^*$ Laboratoire d'Etudes des Propri\'etes Electroniques des Solides,
 Centre National de la Recherche Scientifique,
BP166, 38042 Grenoble, France} 
\address{$^\dagger$ School of Physics and Astronomy, Raymond and Beverly Sackler 
Faculty of Exact Sciences, Tel Aviv University, Ramat Aviv, Tel Aviv 69978, 
Israel}
\maketitle

\begin{abstract}
Approaching a two-component tip made of a superconductor (S) and a
ferromagnet (F) from a magnetic sample allows for two distinct tunneling
processes between the ferromagnets, through S: i) Charge and spin are
conserved; ii) Charge and spin are reversed, e.g. a Cooper pair flows from
S, one electron going into F, the other into the sample.  At subgap
voltages, this allows two currents to flow from the tip : one is
insensitive to the spin polarizations and allows for surface topography,
the other directly tracks the relative spin polarizations of F and the
sample.  The whole device acts as a STM sensitive to the spin polarization
at the Fermi level (MSTM).  Its sensitivity is studied and optimized with
respect to the tip geometry.
\end{abstract}

\bigskip 
\section{Introduction}
Nanoscale characterization of magnetic surfaces is nowadays a major
challenge.  Scanning Tunneling Microscopy (STM) is required, but face the
difficulty of measuring the spin polarization (SP) at the atomic scale, and
simultaneously recording the topographic information like a standard STM.
The simplest (in principle) method uses a ferromagnetic tip, forming a F/I/F 
junction
through vacuum, the tunnel current between two ferromagnets depending on
their relative SP's. This technique has been successfully implemented by
Wiesendanger's group \cite{Wiesen1,Wiesen2}.  Different
proposals rely on ferromagnet-semiconductor junctions \cite{semicon}, or 
spin-orbit coupling in a two-terminal tip \cite{Bruno}.

In a different context, superconducting-ferromagnetic tunnel junctions
(S/I/F) were used by Tedrow and Meserwey \cite{Tedrow} to measure the
exchange field, using the spin splitting of the superconducting density of
states peaks.  For good contacts, Andreev reflection offers an alternative
access to SP : at a normal metal-superconductor (S/N) interface, subgap
conductance is nonzero if an electron (hole) coming from N can be reflected
as a hole (electron) and form a Cooper pair in S \cite{Andreev,BTK}.  This
involves spin reversal in the conventional case of singlet superconductor
(considered here), e.  g. the Andreev current requires electronic channels
for opposite spin close to the Fermi level.  Therefore SP must reduce the
Andreev conductance \cite{deJBeen}.  Measurement of SP through this
reduction was realized by Soulen et al \cite{Soul} and Upadhyay et al. 
\cite{Upad}.  The first experiment uses a superconducting point contact on
a ferromagnetic surface, with a high energy sensitivity at the Fermi level,
below the superconducting gap.  However, contrarily to the spin valve
principle which uses another ferromagnet as a reference and can measure
also the direction of SP, the S/F interface only allows to measure the
absolute value of the polarization, an obstacle against direct domain
imaging.  Moreover, operating the superconducting tip as a STM
rather than a point contact (forming a S/I/F interface through vacuum)
would result in an extremely small Andreev conductance, as it is basically
a {\it two-electron} tunneling process through vacuum.

\section{Proposal for a mixed S/F tip}
We propose here to use a mixed superconductor- ferromagnetic tip, and
combine the energy resolution of the superconductor and the SP direction
sensitivity of the spin-valve effect.  A superconducting tip (S) forms two
interfaces : one with a magnetic layer (denoted as B) coating the tip, except
for its end, and one (vacuum) tunnel interface with the sample (denoted as A)
[Fig.1(a)].  The whole tip (S and B) is biased at the voltage $V < \Delta$,
the superconducting gap.  Some properties of such an F/S/I/F
double interface have been recently studied \cite{us1,us2} : if the
interfaces S/A and S/B are close enough, coherent transfer
of single quasiparticles can occur at both interfaces
\cite{Byers}, and depend on the relative SP's of A and B
\cite{us1,us2,Martin,Choi,Melin}.  For instance, quasiparticles of the same
spin tunnel from B to A through S, and the two resulting currents at the
interfaces S/A and S/B are just opposite, carrying the same SP
[Fig.1(b)].  This "normal" channel, conserving spin and charge, can be
called elastic cotunneling (EC) \cite{AverinNazarov}.  It can be compared 
with usual spin-dependent tunneling at a F/I/F interface, and
is hindered by antiparallel polarizations of A,B. On the contrary, the
latter situation favours an "anomalous" channel, exclusively opened by the
superconductor : a Cooper pair can leave S, made of one quasiparticle going
into A and another into B, with opposite spins.  This mechanism can be also
viewed as tunneling from B to A, but reversing spin and charge : it can be
denoted as crossed Andreev (CA), and generalizes
\cite{Lambert} the usual Andreev process which may take place at each
electrode separately \cite{BTK}.  Both processes EC and CA probe the single
particle propagators (normal for EC, and anomalous for CA) in S, in the
narrow region between the two contacts, and decay over the coherence length
$\xi$ \cite{Byers,Choi,us2}. Since 
$\xi$ can exceed hundredths of nanometers, the range of both tunneling channels 
drastically exceeds that of usual tunnel effect through an insulator. 

To modelize this geometry, one may assume that tunneling occurs at the end
of a nanometric protuberance [Fig.2(a)].  Neglecting curvature effects,
this can be schematized in a planar geometry, with the superconductor
connected to two ferromagnets by one point-like interface A of size $a$
(corresponding to the tip-surface tunnel barrier at the protuberance), and
a larger ring-like interface B, with internal radius $R$ (Fig.2b).  At
radius larger than $R + d$, the F layer is assumed to be isolated from S by
an insulating layer (region B').  The tip contains a few conduction
channels and we assume $a << R$.
 
\begin{figure}
  \centerline{\epsfxsize=8cm
\epsfbox{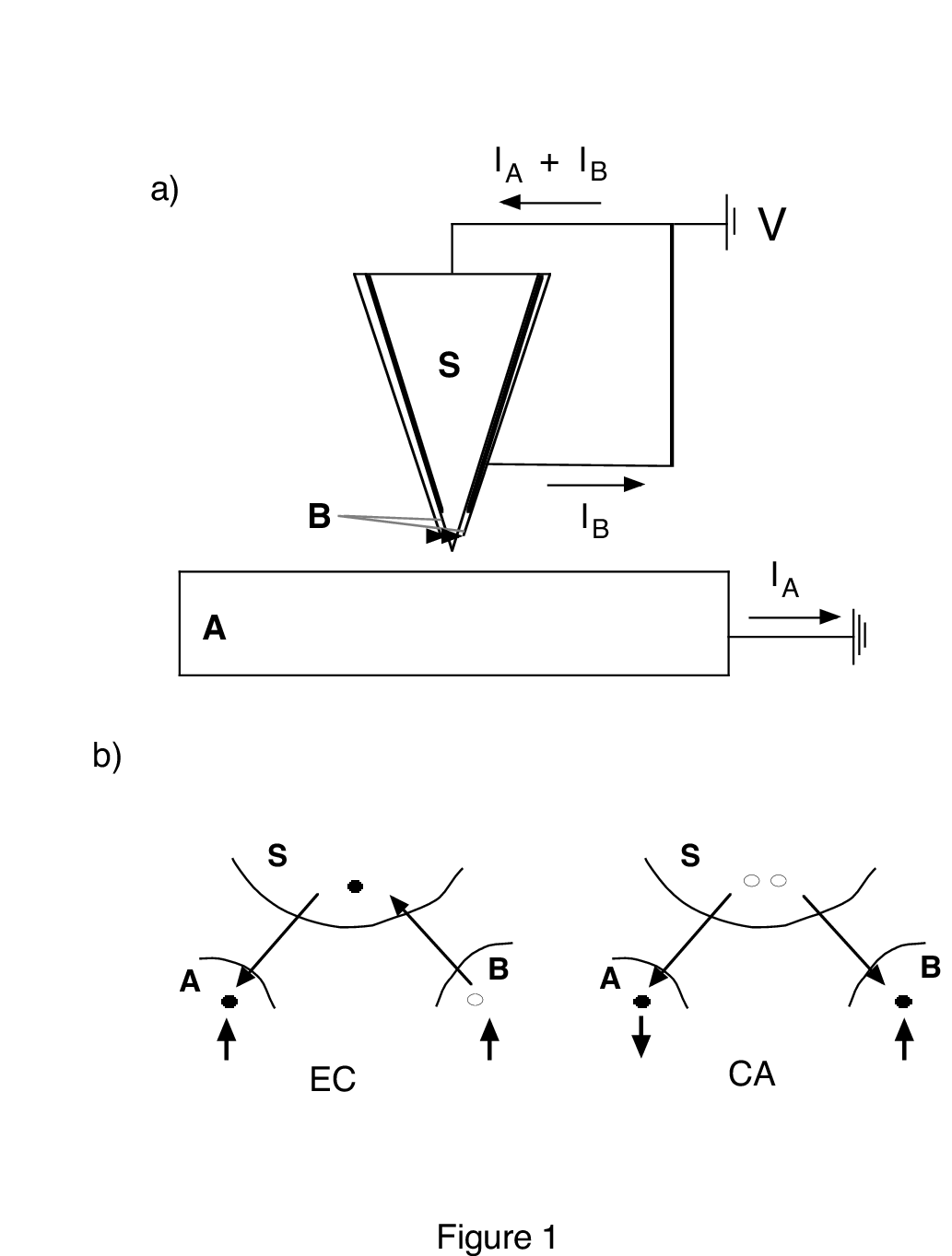}}   
\vspace{4mm}
\caption{{\bf a)} A superconducting tip covered by a ferromagnetic layer 
(B), except for its end, is biased at a voltage $V$ with respect to a 
magnetic sample (A).  {\bf b)}  Sketch of the two basic processes : 
elastic cotunneling (spin-conserving) 
and crossed Andreev, with decay of a Cooper pair in A and B (opposite 
spins)}. 
\end{figure}

\section{Calculation of the conductance matrix}
The currents $I_A$, $I_B$ flowing from S to the ferromagnets A(B) can be
expressed as functions of the external bias $V$.  Notice that a small bias
$\delta V_{B} = V_{S} - V_{B}$ will also be generated at the SB interface,
due to the voltage drop in B induced by $I_{B}$, but as we show later it
can be neglected under specific conditions.  At voltages $V$ smaller
than the gap $\Delta$, and at low temperature, single-particle
contributions are negligible, and $I_{A}$,$I_{B}$ are linear functions of
$V$, as the result of four kinds of two-particle currents \cite{us2} : i)
the Andreev currents at contacts A,B separately, $I_{2A} = G_{2A}V$,
$I_{2B} = G_{2B} \delta V_B$; ii) the crossed Andreev (CA) currents
$I_{CA,A} = I_{CA,B} = G_{CA} (V + \delta V_{B})$ flowing into A and B;
iii) the cotunneling (EC) currents $I_{EC,A} = -I_{EC,B} = G_{EC} (V -
\delta V_{B})$ flowing from B to A (Fig.  1).  As a result, setting
$G_{\pm} = G_{CA} \pm G_{EC}$, one has

\begin{eqnarray}
\label{eq:current}
I_{A}= (G_{2A} + G_{+}) V + G_{-}\, \delta V_{B}\\
I_{B}= G_{-} V + (G_{2B} + G_{+})\, \delta V_{B}
\end{eqnarray}

The single-junction Andreev conductances $G_{2i}$, $i = (A,B)$ are 
given for each spin channel $\sigma$ by \cite{HekkingNazarov} $G_{2i}^{\sigma} \, 
\approx \, {h 
\over e^{2}} \,{\cal G}_{i}^{\sigma}\, {\cal G}_{i}^{-\sigma}\,\, k_{F}^{-2}{\cal S}_{i}
\, \log(k_{F}^{2}{\cal S}_{i})$, 
where $k_{F}$ is the Fermi number in S, ${\cal S}_{i}$ is the area of contact 
$i$  
and the ${\cal G}_{i}^{\sigma}$'s are the single particle conductances per 
unit area for S being in the normal state (Eq. (8)), given by

\begin{equation}
{\cal G}_{i}^{\sigma}  \sim {4\pi e^{2} \over \hbar}  \, 
N_{i}^{\sigma}(0) \, N_{S}(0) \,{ |t_{i}|^2 \over k_{F}^{2}}
\end{equation}

 where $N_{i}^{\sigma}$ is the spin-dependent density of states at 
 the Fermi level in A or B, $N_{S}(0)$ the normal state density of 
 states in S and $t_{i}$ tunneling matrix elements at the interfaces 
 S/A, S/B. $G_{EC}$ and 
$G_{CA}$ have been recently calculated for two separated contacts of size 
much smaller than $\xi$, as a function of the relative position 
$\vec{R}$ of the contacts and their spin polarizations \cite{us2}. 
In  multichannel junctions with normal (non magnetic) 
$A,B$ electrodes one finds 
$G_{CA}=G_{EC}$, thus the crossed conductance $dI_{B}/dV$  vanishes 
\cite{keldysh}.
This symmetry is broken if $A,B$ are spin-polarized ferromagnets and 
in particular the crossed conductance can be either positive or negative. 

The calculation can be performed in the geometry of
Fig.  2b.  The transition rates $\Gamma_{B\to A}^{\sigma}$ and $\Gamma_{S
\to AB}^{\sigma}$ respectively associated to EC and CA processes can be
calculated, using Fermi's golden rule \cite{us2,HekkingNazarov}

\begin{eqnarray}
\label{eq:rate-EC}
\Gamma_{B\to A}^{\sigma} &=& {2 \pi \over \hbar} \;
\int \!\! d\varepsilon d\varepsilon^{\prime} d\zeta d\zeta^{\prime} \;
\delta(\varepsilon- \varepsilon^{\prime}) \;
f(\varepsilon + e\delta V_{B}) \; [1- f(\varepsilon^{\prime}+e V)]
\nonumber
\\&& \hskip40mm
{\mathrm{F}}_{EC}(\zeta, \varepsilon)
\; {\mathrm{F}}_{EC}(\zeta^{\prime}, \varepsilon^{\prime})
\;\;
\Xi_{EC}^{\sigma}(\varepsilon + e\delta V_{B}, \varepsilon^{\prime}+e V
,\zeta,\zeta^{\prime})
\hskip5mm
\end{eqnarray}

and

\begin{eqnarray}
\label{eq:rate-CA}
\Gamma_{S \to AB}^{\sigma} &=& {2 \pi \over \hbar} \;
\int \!\! d\varepsilon d\varepsilon^{\prime} d\zeta d\zeta^{\prime} \;
\delta(\varepsilon + \varepsilon^{\prime}) \;
f(\varepsilon + e\delta V_{B}) \, {f}(\varepsilon^{\prime}+e V)
\nonumber
\\&& \hskip40mm
{\mathrm{F}}_{CA}(\zeta, \varepsilon)
\; {\mathrm{F}}_{CA}(\zeta^{\prime}, \varepsilon^{\prime})
\;\;
\Xi_{CA}^{\sigma}(\varepsilon + e\delta V_{B}, \varepsilon^{\prime}+eV
,\zeta,\zeta^{\prime})
\hskip5mm
\end{eqnarray}

where $f(\varepsilon)$ is the Fermi function,
${\mathrm{F}}_{EC}(\zeta, \varepsilon) \;=\;
(\zeta + \varepsilon)/ (\zeta^2 + \Delta^2 - \varepsilon^2)$
and
${\mathrm{F}}_{CA}(\zeta, \varepsilon) \;=\;
\Delta/ (\zeta^2 + \Delta^2 - \varepsilon^2)$. Quasiparticle propagation
in the specific geometry is described by the functions
$\Xi(\varepsilon, \varepsilon^{\prime},\zeta,\zeta^{\prime})$. 
Assuming planar uniform tunnel junctions, local tunneling, 
$t(\vec{r},\vec{r^{\prime}}) \,=\, t  \,
\delta (\vec{r}-\vec{r^{\prime}})$ and ballistic
propagation in S, A and B for simplicity, both functions $\Xi_{EC}$
and $\Xi_{CA}$ can be expressed as
\begin{eqnarray}
\label{eq:Xi-funct}
\Xi^{\sigma}(\varepsilon, \varepsilon^{\prime},\zeta,\zeta^{\prime}) &=&
|t_A t_B|^2 \int_{A} \hskip-1mm d\vec{r}_1 d\vec{r}_2 \;
\int_{B} \hskip-1mm d\vec{r}_3 d\vec{r}_4  \; \;
\rm{J}_{A}^{\sigma}(12,\varepsilon)\,
\,\rm{J}_S^{\sigma}(31,\zeta)\,
\rm{J}_S^{\sigma}(24,\zeta^{\prime})\,
\rm{J}_{B}^{\pm\sigma}(43, \varepsilon^{\prime})
\end{eqnarray}
where $+\sigma$ ($-\sigma$) in $\rm{J}_{B}^{\pm\sigma}$ applies for EC (CA)
and the spectral functions are defined as
$\rm{J}_A^{\sigma}(12, \omega) \equiv
\rm{J}_A^{\sigma}(\vec{r}_1,\vec{r}_2, \omega)
\,=\, \sum_k \delta(\omega-\varepsilon_{\mathbf{k}{\sigma}}) \,
\psi_{k\sigma}(\vec{r}_1) \psi_{k\sigma}^*(\vec{r}_2)$.
The
integrals in (\ref{eq:Xi-funct}) run on the contact surfaces.

At low temperature and voltages, a low-energy expansion yields the 
cotunneling and CA currents
$I_{EC}^{\sigma} = e \Gamma_{B\to A}^{\sigma}$
and
$I_{CA}^{\sigma} = e \Gamma_{S \to AB}^{\sigma}$.
Let us for simplicity assume equality of the Fermi wavevectors on each 
sides of the interfaces. This 
allows to write the two-electron conductances $G_{EC}^{\sigma}$ and 
$G_{CA}^{\sigma}$ for spin $\sigma$ as

\begin{eqnarray}
\label{EC-CA-conductance}
G_{EC}^{\sigma} \approx\;{\pi^{2} h \over 16 e^2} \; {\cal G}_A^{\sigma}\,{\cal G}_{B}^{\sigma}\;
{{\cal S}_{A}\over k_{F}^{2}}\, f({R \over \xi},{d \over \xi}) \\
\bigskip
G_{CA}^{\sigma} \approx\;{\pi^{2} h \over 16 e^2} \; {\cal G}_A^{\sigma}\,{\cal G}_{B}^{-\sigma}\;
{{\cal S}_{A}\over k_{F}^{2}}\, f({R \over \xi},{d \over \xi})
\end{eqnarray}

\noindent
which have the {\it same} geometrical dependence, and only differ through 
the spin-dependent conductances ${\cal G}_i^{\sigma}$.
 In the ring contact 
geometry (Fig. 2b), the dependence on the internal 
radius $R$ and the width $d$ of contact B is 
determined by  $f({R \over \xi},{d \over \xi}) = 
\int_{2R/\xi}^{2(R + d)/\xi} e^{-x}dx/x$, thus $G_{CA},G_{EC}$ vanish 
for $R >> \xi$. If $R < \xi << d$, $f({R \over \xi},{d \over \xi}) 
\approx \log(\xi /2R)$. And in the case $R,d < \xi$, one finds 
 $f({R \over \xi},{d \over \xi}) 
\approx \log( 1 + {d\over R})$.

\begin{figure}
  \centerline{\epsfxsize=8cm
 \epsfbox{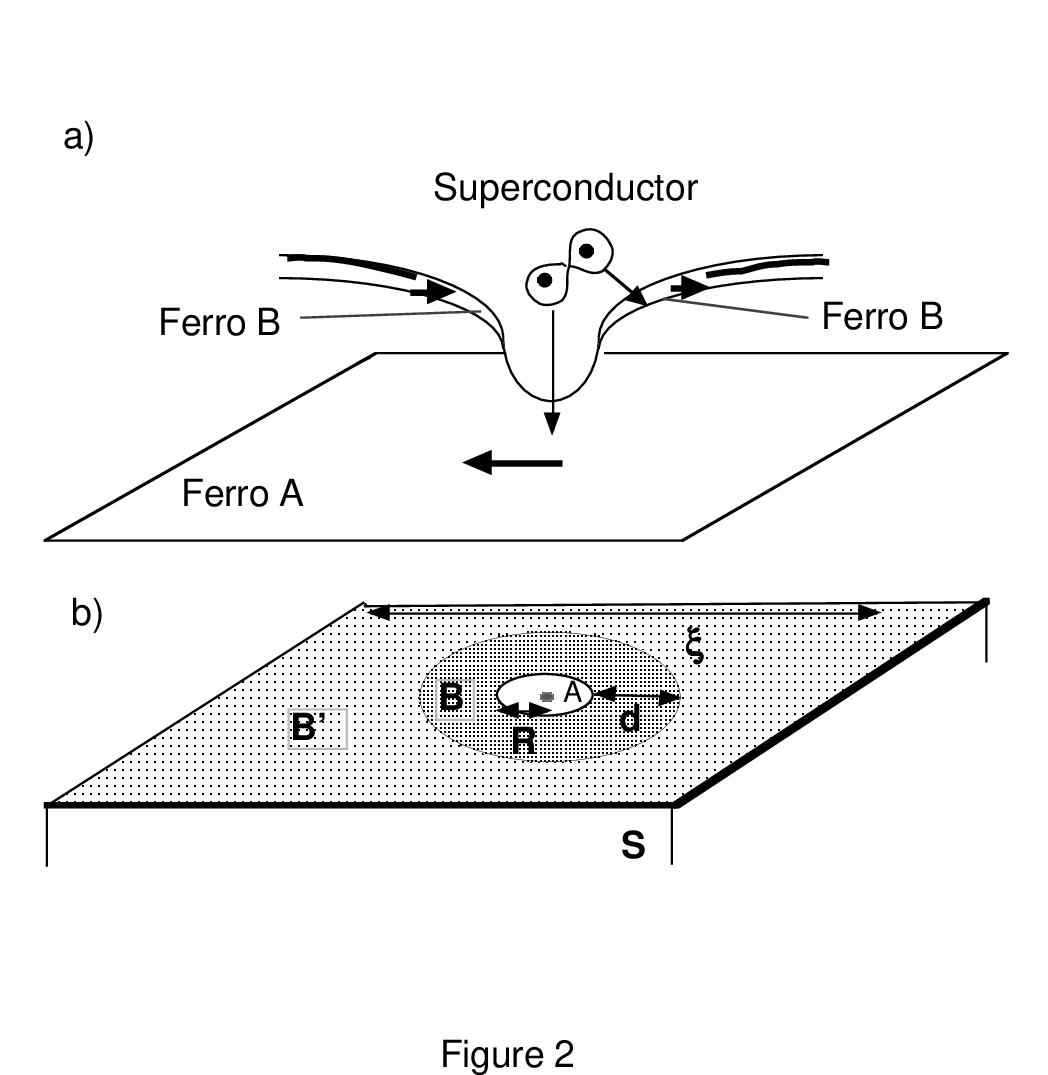}}   
  \vspace{4mm}
  \caption{{\bf a)} Enlarged view of the tip end, where one of the crossed 
  process (CA) is represented. {\bf b)} Two-dimensional modelization 
  of the interfaces S/A (small region A) and S/B (shaded area). The 
  region B' is isolated from S (see text)}.
 \end{figure}

\section{Discussion}
The logarithmic dependence of the crossed conductances $G_{\pm}$ with the
size of B has important consequences in terms of the sensitivity of the
proposed device.  Let us discuss the various contributions in Eqs.  1,2. 
First, if the S/A interface is tunnel-like and if S/B is good, $t_{A} <<
t_{B}$ therefore $G_{2A} / G_{\pm} \approx {{\cal G}_{A} \over {\cal
G}_{B}} << 1$, showing that crossed processes dominate over Andreev
processes at the tip end (two-electron tunneling into A).  On the other
hand, $G_{2B} / G_{\pm} \approx {{\cal G}_{B} {\cal S}_{B} \over {\cal
G}_{A} {\cal S}_{A}} >> 1$.  Yet, one can verify that the contributions in
$I_{A,B}$ due to $\delta V_{B}$ are negligible in practice.  In fact,
$\delta V_{B} \sim -R_{B} I_{B}$ where $R_{B}$ is the ferromagnetic thin
film resistance, thus $I_{B} = {G_{-} \over (1 + G_{2B}R_{B})} V$.  And
$G_{2B}R_{B}
\approx ({e^{2} \over h} R_{B}) {1 \over Z_{B}^{4}}N_B$
 where $Z_{B} \sim {\epsilon_{F} \over t_{B}}$ can be identified 
 \cite{Cuevas} with the BTK parameter for 
 the S/B interface \cite{BTK}, $N_B \approx k_F^2{\cal S}_{B}$ being the 
 number of conduction channels in B . The potential drop $\delta V_{B}$ can be 
 safely neglected if $R_B \, Z_{B}^{-4} \, (k_{F}^{2}{\cal 
 S}_{B}) << 10^{4}$ ($R_B$ expressed in Ohms) which can be reasonably 
 fulfilled, since the area of contact B can be reduced without affecting much the 
 crossed conductances $G_{\pm}$ which vary logarithmically with $d/R$ 
 (for instance $R_{B} <  10 \Omega$, $Z_{B}^2 \sim 10$ and $k_{F}^{2}{\cal 
 S}_{B} \sim 10^{3}$). 
 
In these conditions, one has $I_{A} \approx G_{+}V$ and $I_{B} \approx
G_{-}V$.  Let us briefly explain the physical mechanism allowing a current
to pass through the S/B interface in absence of a bias applied directly at
this interface: the tip bias $V$ at the S/A interface forces a
quasiparticle to tunnel from S to A under the condition that another
quasiparticle is pulled simultaneously through S/B, due to the EC and CA
processes described above.  Either process increase the effective
conductance $I_{A}/V$ at contact A. Conversely, the EC and CA contributions
to the current induced at S/B have {\it opposite signs} [Fig.1(b)].
 
Let us now examine the spin sensitivity of the proposed device. 
Eqs. (7,8) show that for a given spin the processes EC and CA have identical 
rates, except that EC connects channels with the same spins in A 
and B while CA connects channels with opposite spins. 
Still assuming the equality of Fermi vectors on both sides of the 
interfaces, let $P_{A,B} =
\frac{N_{A,B}^{\sigma} - N_{A,B}^{-\sigma}}{N_{A,B}^{\sigma} +
N_{A,B}^{-\sigma}}$ be the spin polarizations in the densities of states at
the Fermi level.  Then one verifies that $G_{EC}$ is proportionnal to
$(1+P_{A}P_{B})$ while $G_{CA}$ is proportionnal to $(1-P_{A}P_{B})$.  It
follows that $G_{+}$, therefore $I_{A}$, is independent on SP, while
$G_{-}$, thus $I_{B}$, is proportionnal to $(-P_{A}P_{B})$, and allows the
comparison of the SP's of A and B. The sign and amplitude of $I_{B}$ will
reflect the local surface SP of A. As a consequence, and this is the
central result of this Letter, measuring simultaneously the currents
$I_{A}$ and $I_{B}$ allows:

i) To operate as a STM, since the current $I_{A}$ is spin-independent 
thus permits the topographic imaging of the surface at the atomic scale. 

ii) To measure the local spin polarization and image the domain structure
of surface A.

Notice that the spatial resolution is here by no means limited 
by $\xi$, but instead  by the atomic-like scale $a$, like an usual STM. 
The response of the device is governed by the crossed conductances 
$G_{\pm}$. 
Taking $ G_{A} \sim 10^{-7}$, $V \sim 10^{-4} V$ (typically one tenth of the gap 
for Niobium) and $Z_{B}^2 \sim 10$ leads to 
currents $I_{A,B}$ of the order of $1 pA$, while the direct Andreev 
current between the tip and the surface is about $100$ times smaller thus 
hardly observable. 

Nevertheless, using a good interface B to enhance the subgap current 
has a drawback : although the direct Andreev current $I_{2B}$ in B is 
negligible on average, its fluctuations dominate 
the Johnson-Nyquist noise, given respectively in A and B (the various 
contributions $I_{2A}$, $I_{2B}$, $I_{CA}$, $I_{EC}$ are uncorrelated) by
$S_A\,\approx 4k_BT(G_{2A} + G_{CA} + G_{EC}) 
\, \approx 4k_BT \, G_{\pm}$ and $
S_B \approx 4k_BT(G_{2B} + G_{CA} + G_{EC}) \,\approx 4k_BT\,G_{2B}$. 
Notice that it is essential here to avoid conductance noise across the
region of the ferromagnetic layer situated at a distance larger than $\xi$
from A. Therefore region B' on Figure 2 must be isolated from S by an
insulating layer.  An estimate of the largest noise contribution gives
${\delta I_B \over I_B} \approx {10^{-9} \sqrt{T} \over V}\,{e^{2} \over h
G_{A}}\,{k_{F}^{2}{\cal S}_{B} \log(k_{F}^{2}{\cal S}_{B}) \over
\log(1+d/R)} / \sqrt {Hz}$ ($T$ in Kelvins, V in volts).  One finds an
optimal noise/signal ratio when $R \sim d $ and $a < {R,d} < \xi$.  With
$G_{A} \sim 10^{-7}S$, $V \sim 10^{-4}V$ and $T \sim 0.1 K$, it is of the
order of $10^{-2} \sqrt{N_B}$ at $10 Hz$.  This implies that the dimensions
of the ring contact B must be nanometric, in order to carry not more than a
few hundred channels.  For a given geometry, a better sensitivity is
realized if the superconductor has a low density of carriers.  Let us
remark here that the ring geometry achieves a major improvement with
respect to two point contacts of sizes $a << \xi$, distant by $R$
\cite{us2,Choi}, 
where the conductance drops by a factor $(k_F\,R)^2$ : in the ring 
geometry presented in this Letter the 
measured currents are higher by a factor $\sim (R/a)^{2}$ while the 
signal/noise ratio is improved by a factor $R/a$, ranging from $10$ to 
$100$ depending on the chosen materials.  

\section{Conclusion}
In summary, we have shown how a new principle for a magnetic 
STM results from non-local two-particle tunneling 
processes at two S/F interfaces. Let us discuss the validity of our simplifying 
assumptions. First, beyond the 
simple ballistic regime considered here, in a dirty superconductor the 
crossed processes  will decay on $\xi \approx 
\sqrt{\xi_{0}l}$ where $l$ is the mean-free path and $\xi_{0}$ 
the BCS coherence length ($l < \xi_{0}$), as EC and CA processes directly
probe the one-particle correlation functions in the superconductor. 
Secondly, proximity effect should be considered : due to the ferromagnet B,
the gap function might be reduced in the superconductor.  Recent
theoretical results \cite{Fazio,Valls} for a
clean interface do not show dramatic effects, which should be even smaller
for an imperfect interface S/B (we assumed above a value $Z^2
\sim 10$).  One might also
worry about vortices induced by the stray field from B. If created, vortex
cores, of size $\xi$, should penetrate in the bulk and hardly affect the
gap in the tunneling region, a priori smaller.  More directly, experiments reported in
\cite{Soul,Upad} demonstrate that the presence of a ferromagnet does not
destroy the Andreev reflection, even with a superconducting tip pressed on
a bulk ferromagnet.  We thus believe that the present proposal could be
realized if choosing properly the material and geometry parameters.  Low
temperature superconductors, with a large coherence length and possibly a
low carrier density are preferable.  As an S/F couple, one may try $Nb/Fe$,
or $Al/Fe$.  One also notices that the interactions between
ferromagnets B and A due to stray field effects should be much less serious
when they sit at a typical distance $\xi$ which can exceed hundredths of
nanometers.  In a configuration of parallel magnetization especially, this
can be a sensible improvement with respect to the "all ferromagnet" MSTM
principle where the AB distance is a few Angstroems.  This is allowed by
the coherence in the superconductor, allowing propagation of quasiparticles
on distances much larger than the width of a usual tunnel barrier.

\bigskip

Acknowledgements: The authors have benefitted from stimulating discussions with H. 
Courtois, P. Mallet, R. M\'elin, P. Nozi\`eres and J. Y. Veuillen. 

\end{document}